\documentclass[aps,prb,preprint]{revtex4-1}%
\usepackage{amsfonts}
\usepackage{amsmath}
\usepackage{amssymb}
\usepackage{graphicx}
\makeatletter

\renewcommand{\p@subsection}{}

\renewcommand{\p@subsubsection}{}

\makeatother
\def\figww{3.8cm}
\def\figw{5cm}
\begin{document}

\title{Charge density waves as the origin of dip-hump structures in differential
tunneling conductance of cuprates: the case of $d$-wave superconductivity}

\author{Alexander M.~Gabovich}
\email{gabovich@iop.kiev.ua}
\affiliation{Institute of Physics, Nauka Ave. 46, Kyiv 03680,
Ukraine}
\author{Alexander I.~Voitenko}
\email{voitenko@iop.kiev.ua}
\affiliation{Institute of Physics, Nauka Ave. 46, Kyiv 03680,
Ukraine}
\date{\today}

\begin{abstract}

Conductance-voltage characteristics (CVCs) for non-symmetric tunnel junctions
between $d$-wave superconductors with charge-density waves (CDWs) and normal
metals were calculated. It was shown that they have a V-like form at small
voltages $V$ and are asymmetric at larger $V$ owing to the presence of CDW
peak in one of the $V$-branches. The spatial scatter of the dielectric (CDW)
order parameter smears the CDW peak into a hump and induces a peak-dip-hump
structure (PDHS) typical of CVCs observed for such junctions. At temperatures
larger than the superconducting critical temperature, the PDHS evolves into a
pseudogap depression. The results agree well with the scanning tunneling
microscopy data for Bi$_{2}$Sr$_{2}$CaCu$_{2}$O$_{8+\delta}$ and YBa$_{2}%
$Cu$_{3}$O$_{7-\delta}$.

\end{abstract}

%insert suggested PACS numbers in braces on next line


%insert suggested keywords - APS authors don't need to do this
%\keywords{}

\maketitle

High-$T_{c}$ superconductors demonstrate a number of features which are widely
discussed but badly understood. Among the most enigmatic of them is the
pseudogap (PG) coexisting with superconductivity at various low and
intermediate dopings.\cite{gabovich10:681070,ekino11:699,sacuto13:022502} In
particular, the differential current-voltage characteristics (CVCs)
$G(V)=dJ/dV$, where $J$ is the quasiparticle current and $V$ the bias voltage,
of tunnel junctions involving high-$T_{c}$ oxides reveal a depletion at
temperatures, $T$, above and below the critical temperature, $T_{c}$. Another
CVC\ peculiarity is a robust peak-dip-hump structure (PDHS) observed at low
temperatures $T$%
,\cite{fischer07:353,ekino07:180503,mourachkine07:956,zasadzinski08:833,krasnov09:214510,suzuki12:010112,berthod13:014528}
with the CVCs in the case of superconductor~(S)--insulator~(I)--normal
metal~(N) junction often demonstrating bias-voltage asymmetry. The PDHS is
also found in the angle-resolved photoemission (ARPES) spectra for a good many
oxides.\cite{kordyuk02:077003,borisenko03:207001,wei08:097005} The existing
explanations of PDHS are mostly based on the assumption of extremely strong
coupling between electrons and either spin
fluctuations\cite{chubukov98:4716,eschrig00:3261} or
phonons.\cite{kulic05:092505,citro06:014527} However, in this case the PG
should be attributed to some other physical reasons, which seems quite
unnatural in view of the conclusions drawn from tunneling spectroscopic
data.\cite{bae08:094519,vedeneev10:054501} Anyway, the PG and the PDHS are
considered as unrelated phenomena.

We propose a different model, which enables all those peculiarities to be
described from the same position. It is based on two pieces of evidence that
can be regarded reliably established for high-$T_{c}$ oxides: (i)~plenty of
cuprates reveal, directly or indirectly, charge density waves (CDWs) competing
with the Cooper pairing-induced reconstruction of the electron
spectrum,\cite{gabovich10:681070,ekino11:699,gabovich13:301} and
(ii)~high-$T_{c}$ superconductors are inherently non-uniform
objects.\cite{boyer07:802,pasupathy08:196,alldredge13:104520} The former is
most probably a consequence of the reduced system dimensionality and the
resulting Fermi surface (FS) nesting,\ whereas the latter may be an intrinsic
feature associated with the oxygen non-stoichiometry.

A characteristic feature of CVCs in the case of non-symmetric S--I--N
junctions is their non-symmetric
behavior.\cite{renner98:149,fischer07:353,zasadzinski08:833,berthod13:014528}
For instance, the PDHS below $T_{c}$ is the most pronounced at a negative bias
voltage polarity, which corresponds to electron ejection, and the pseudogap
depletion above $T_{c}$ is also non-symmetric with respect to the bias
polarity. Moreover, the PDHSs are sometimes observed in both branches, the
ratio between their magnitudes being different. Such a non-symmetricity can be
easily interpreted in the framework of our theory by making allowance for the
CDW phase (see below). Nevertheless, the adopted phenomenological approach
makes the appearance of PDHS in either of the branches
equiprobable.\ Therefore, an additional microscopic consideration should be
invoked for the explanation of CVC non-symmetricity.

We should mention three additional scenarios ensuring the CVC asymmetry. One
of them involves a substantial role of the Van Hove singularity in the density
of states.\cite{levydecastro08:267004,levydecastro10:099702,berthod13:014528}
The acceptance of this viewpoint leads to certain
problems,\cite{onufrieva10:099701,carbotte11:066501} the discussion of which
goes beyond the scope of this article. The second one suggests the decisive
role of the superconducting-gap energy dependence $\Delta(E)$, so that the
slope $d\Delta(E\simeq E_{F})dE$ is responsible for the
asymmetry.\cite{hirsch99:11962} Here, $E_{F}$ is the Fermi energy. However, in
this case neither of the two humps is suppressed, which contradicts the
observations. The third scenario introduces strong many-body correlations
making electron- and hole-like excitations nonequivalent.\cite{anderson06:1}
The importance of the factor concerned cannot be ruled out, in principle,
although the absence of the gap-driven peak asymmetry in other strongly
correlated superconductors does not count in favor of this viewpoint.

In this article, we restrict the consideration to tunnel spectra measured for
non-symmetric S-I-N junctions with the quasiparticle current flowing along the
crystal $c$-axis, i.e. perpendicularly to CuO$_{2}$ layers, which as
appropriate to the scanning tunneling microscopy (STM) setup. Moreover, we
consider only the case of $d_{x^{2}-y^{2}}$-wave symmetry of the
superconducting order parameter predominately adopted by the
community,\cite{tsuei08:869} although its true symmetry is still not
known.\cite{klemm05:801,zhao11:038302}

Our model of the partially gapped CDW $d$-wave superconductor (CDWS) with the
two-dimensional electron
spectrum\cite{gabovich10:681070,ekino11:699,gabovich13:104503,gabovich13:301}%
\ was developed on the basis of its predecessor developed for the partially
gapped (dielectrized) $s$-wave
superconductor.\cite{bilbro76:1887,gabovich03:2745} It was applied to cuprates
with the checkerboard (biaxial, the number of the CDW sectors $N=4$) or
unidirectional ($N=2$) CDW patterns. Here, for brevity, we present only the
results obtained for $N=4$. The second electrode\textrm{ }is chosen to be a
normal metal with a constant\ electron density of states; e.g., a tip of the
STM device.

\begin{figure}[tb]
\includegraphics[width=\figww]{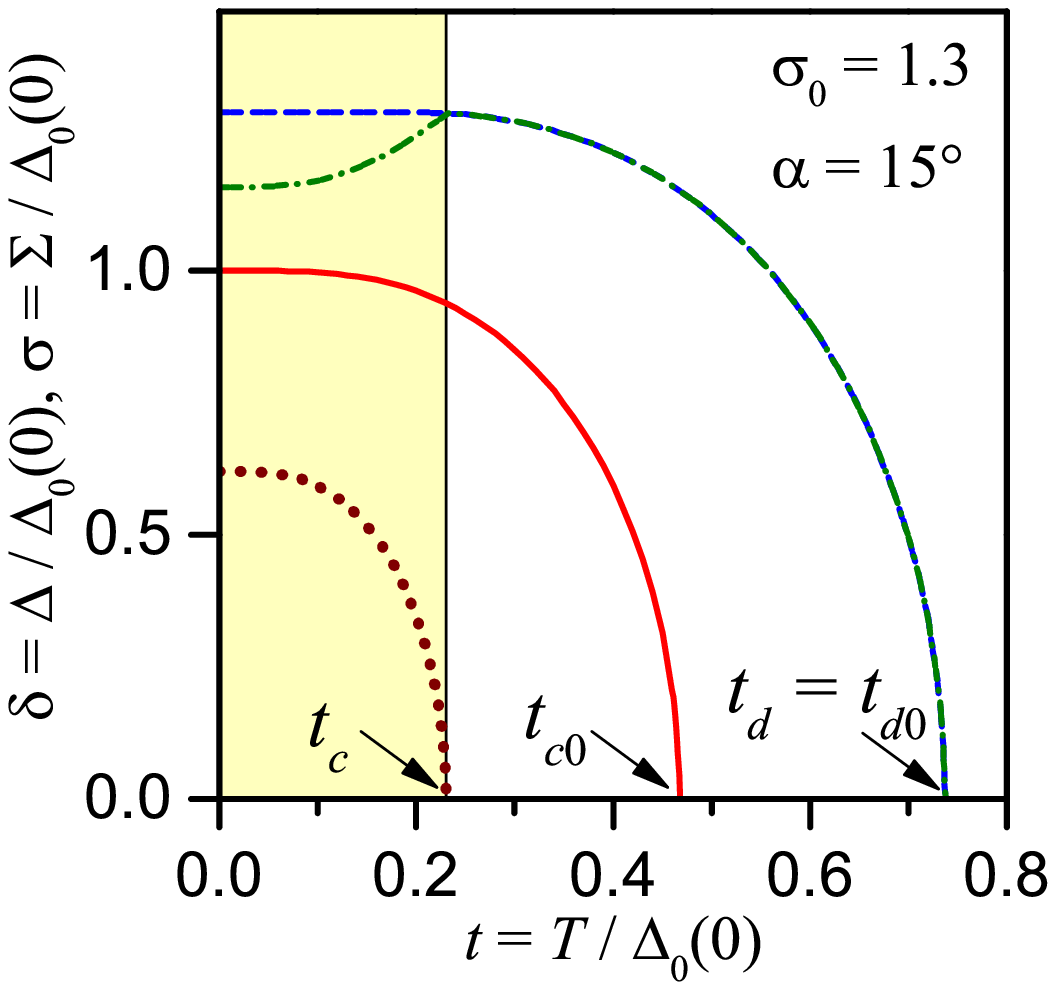}(a)
\includegraphics[width=\figww]{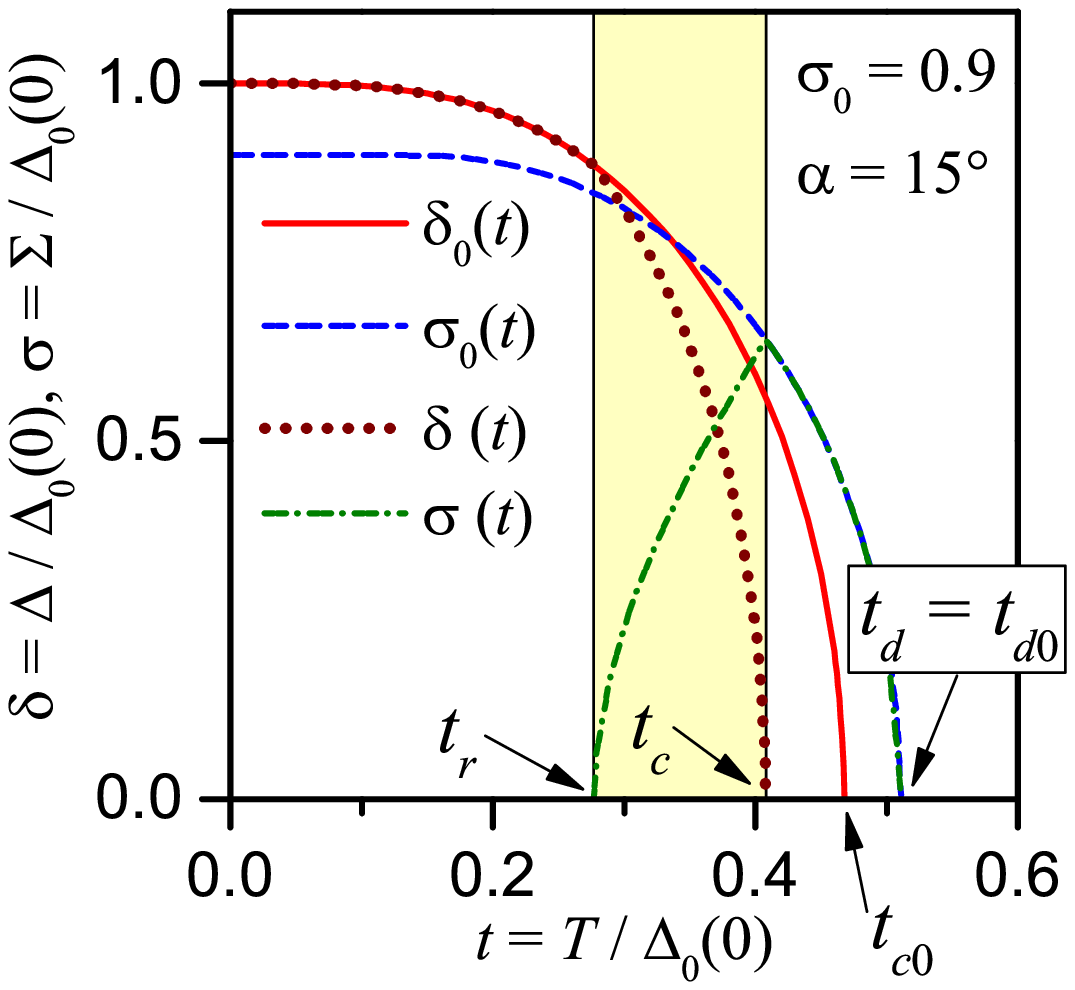}(b) \caption{(Color online)
Dependences of the dimensionless parent (subscript~0, in the absence of
competing pairing) and actual superconducting, $\delta$, and charge density
wave (CDW), $\sigma$, order parameters on the normalized temperature,
$t=T/\Delta_{0}(0)$ without (panel a) and with (panel b) the CDW reentrance.
$\delta(t)=\Delta(T)/\Delta_{0}(0)$, $\sigma_{(0)}(t)=\Sigma_{(0)}%
(T)/\Delta_{0}(0)$. The critical temperatures appropriate to the parent states
($t_{c0}$ and $t_{d0}$) and actual ($t_{c}$, $t_{d}$, and $t_{r}$)
CDW\ superconductors are indicated by arrows. The temperature regions where
the CDWs and superconductivity coexist ($\sigma\neq0$ and $\delta\neq0$) are
painted. The corresponding values of $\sigma_{0}$ and $\alpha$, the half-width
of the CDW sectors, are specified.}%
\label{fig1}%
\end{figure}

The $d$-wave CDWS spectrum is a result of interaction between two pairing
mechanisms, the non-isotropic net electron-electron attraction and the
isotropic electron-hole (excitonic or Peierls) one. When the Cooper pairing is
\textquotedblleft switched off\textquotedblright, only a parent CDW phase with
the complex zero-temperature dielectric order parameter $\Sigma_{0}%
(0)e^{i\varphi}$ and the critical temperature of partial dielectrization
$T_{d0}=\frac{\gamma}{\pi}\Sigma_{0}(0)$ exists. Here, $\gamma=1.78\ldots$ is
the Euler constant, and the Boltzmann constant $k_{B}=1$. At $T<T_{d0}$, the
magnitude of dielectric order parameter varies as
\begin{equation}
\Sigma_{0}(T)=\Sigma_{0}(0)\mathrm{M\ddot{u}}_{s}(T/T_{d0}),
\end{equation}
where $\mathrm{M\ddot{u}}_{s}(x)$ is the standard ($s$-wave) M\"{u}hlschlegel
dependence with $\mathrm{M\ddot{u}}_{s}(0)=1$
(Ref.~\onlinecite{muhlschlegel59:313}, see Fig.~\ref{fig1}). The function
$\Sigma_{0}(T)$ is assumed to be constant (the $s$-wave symmetry) within each
of four FS d-sectors. These sectors are oriented crosswise in pairs along the
lattice $\mathbf{k}_{x}$ and $\mathbf{k}_{y}$ axes in the momentum space, with
each of the sectors having the angular width $2\alpha$. Introducing the
angular\ factor $f_{\Sigma}(\theta)$ equal 1 within each sector and 0 outside,
where $\theta$ is the angle in the two-dimensional plane in the momentum space
reckoned, e.g., from the $\mathbf{k}_{x}$ direction, the profile of the parent
dielectric order parameter over the FS can be presented in the
factorized\ form
\begin{equation}
\bar{\Sigma}_{0}(T,\theta)=\Sigma_{0}(T)f_{\Sigma}(\theta).
\end{equation}
At the same time, if the CDW pairing is \textquotedblleft switched
off\textquotedblright, we obtain a parent $d$-wave BCS
superconductor\cite{won94:1397} with the lobes of the superconducting order
parameter $\bar{\Delta}_{0}(T,\theta)$ also oriented in the $\mathbf{k}_{x}$
and $\mathbf{k}_{y}$ directions, i.e. in the same (antinodal) directions as
the bisectrices of CDW sectors (the $d_{x^{2}-y^{2}}$-wave symmetry). In
contrast to the parent $\Sigma$ order parameter, the $\mathbf{k}$-space
profile of its $\Delta$-counterpart is extended over the whole FS. It can also
be presented in the factorized form
\begin{equation}
\bar{\Delta}_{0}(T,\theta)=\Delta_{0}(T)f_{\Delta}(\theta) \label{Delta0}%
\end{equation}
with the angular factor $f_{\Delta}(\theta)=\cos2\theta$. At $T<T_{c0}%
=\frac{\gamma\sqrt{\tilde{e}}}{2\pi}\Delta_{0}(0)$, where $\tilde{e}$ is the
base of natural logarithms,
\begin{equation}
\Delta_{0}(T)=\Delta_{0}(0)\mathrm{M\ddot{u}}_{d}(T/T_{c0}),
\end{equation}
where $\mathrm{M\ddot{u}}_{d}(x)$ is the superconducting order parameter
dependence in the case of $d$-wave pairing.\cite{won94:1397}

While describing the interplay between both pairings, we assume the angular
functions $f_{\Sigma,\Delta}(\theta)$ to remain intact. This mutually
detrimental interplay leads to the drastic difference of the actual
$\Sigma(T)$ and $\Delta(T)$ functions from the parent ones---$\Sigma_{0}(T)$
and $\Delta_{0}(T)$, respectively---in the interval of their coexistence [see
Fig.~1(a)]. The corresponding profile [cf. Eq.~(\ref{Delta0})]
\begin{equation}
\bar{\Delta}(T,\theta)=\Delta(T)f_{\Delta}(\theta)
\end{equation}
emerges on the non-dielectrized (nd) FS sections, and the gap
\begin{equation}
\bar{D}(T,\theta)=\sqrt{\bar{\Sigma}^{2}(T,\theta)+\bar{\Delta}^{2}(T,\theta
)},
\end{equation}
where
\begin{equation}
\bar{\Sigma}(T,\theta)=\Sigma(T)f_{\Sigma}(\theta),
\end{equation}
on the dielectrized (d) ones. The relevant self-consistent set of equations,
which is to be solved to determine $\Sigma(T)$ and $\Delta(T)$ for the given
set of problems parameters $\left[  \Delta_{0}(0),\Sigma_{0}(0),\alpha\right]
$, can be found
elsewhere.\cite{gabovich09:224501,gabovich10:681070,ekino11:699,ekino11:385701}
Now, neither of the order parameters can be described by the function
$\mathrm{M\ddot{u}}_{s}(T/T_{c})$ or $\mathrm{M\ddot{u}}_{d}(T/T_{d})$.
Besides, one of the parent critical temperatures, $T_{c0}$ or $T_{d0}$---to be
more accurate, the minimum one---changes to $T_{c}$ or $T_{d}$, respectively.
Moreover, at some $\left[  \Delta_{0}(0),\Sigma_{0}(0),\alpha\right]
$-combinations, the phenomenon of $\Sigma(T)$-reentrance emerges consisting in
that $\Sigma(T)\neq0$ within a certain temperature interval $0<T_{r}<T<T_{d}$ [Fig.~1(b)].

The quasiparticle current $J(V)$ flowing through an CDWS-I-N junction along
the crystal $c$-axis includes three components,
\begin{equation}
J(V)=\frac{1}{2\pi}%
%TCIMACRO{\dint \limits_{-\pi}^{\pi}}%
%BeginExpansion
{\displaystyle\int\limits_{-\pi}^{\pi}}
%EndExpansion
d\theta\left[  J_{n}\left(  V\right)  +J_{d}\left(  V\right)  +J_{c}\left(
V\right)  \right]  . \label{Jns}%
\end{equation}

Here,%
\begin{equation}
J_{n}=\frac{1-f_{\Sigma}(\theta)}{4eR}%
%TCIMACRO{\dint \limits_{-\infty}^{\infty}}%
%BeginExpansion
{\displaystyle\int\limits_{-\infty}^{\infty}}
%EndExpansion
d\omega~K(\omega,V,T)\left\vert \omega\right\vert f(\omega,\bar{\Delta}),
\end{equation}%
\begin{equation}
J_{d}=\frac{f_{\Sigma}(\theta)}{4eR}%
%TCIMACRO{\dint \limits_{-\infty}^{\infty}}%
%BeginExpansion
{\displaystyle\int\limits_{-\infty}^{\infty}}
%EndExpansion
d\omega~K(\omega,V,T)\left\vert \omega\right\vert f(\omega,\bar{D}),
\end{equation}%
\begin{align}
J_{c}  &  =\frac{f_{\Sigma}(\theta)\Sigma\left(  T\right)  \cos\varphi}%
{4eR}\nonumber\\
&  \times%
%TCIMACRO{\dint \limits_{-\infty}^{\infty}}%
%BeginExpansion
{\displaystyle\int\limits_{-\infty}^{\infty}}
%EndExpansion
d\omega~K(\omega,V,T)~\mathrm{sign}\left(  \omega\right)  f(\omega,\bar{D}),
\label{jc}%
\end{align}
the factor%
\begin{equation}
f(\omega,x)=\frac{\theta\left(  \left\vert \omega\right\vert -x\right)
}{\sqrt{\omega^{2}-x^{2}}}%
\end{equation}
is associated with the density of states in the gapped CDWS, the kernel%
\begin{equation}
K(\omega,V,T)=\tanh\frac{\omega}{2T}-\tanh\frac{\omega-eV}{2T}%
\end{equation}
makes allowance for the Fermi statistics of charge distribution over the
energy levels, $e>0$ is the elementary charge, and $R$ is the normal-state
resistance of the junction. Any issues related to the tunneling directionality
are left beyond the scope of consideration, since they lead only to minor
quantitative corrections. Term~(\ref{jc}) is generated by Green's function
describing the electron-hole dielectric
pairing.\cite{gabovich10:681070,ekino11:699,gabovich13:301} The CDW phase
$\varphi$ entering the expression for the Green's function is usually pinned
by the junction interface and acquires the values 0 or $\pi$. This
circumstance is responsible for the CVC asymmetry\cite{gabovich97:7785} (see
also the earlier work, Ref.~\onlinecite{artemenko84:691}) needed to reproduce
$G(V)$'s observed for non-symmetric junctions with high-$T_{c}$
superconductors.\cite{fischer07:353,zasadzinski08:833,jenkins09:227001}

The further consideration is convenient to be carried out using the normalized
quantities $\sigma_{0}=\Sigma_{0}(0)/\Delta_{0}(0)$, $t=T/\Delta_{0}(0)$,
$v=eV/\Delta_{0}(0)$, and $j=\frac{4eR}{\Delta_{0}(0)}J$. The dimensionless
conductance $g(v)=R\frac{dJ}{dV}$ was found by reproducing the actual
procedure of experimental $G(V)$ determination (see discussion in
Ref.~\onlinecite{ekino08:425218}); namely, we numerically calculated the ratio%
\begin{equation}
g(v)\approx\frac{j(v+\delta v)-j(v-\delta v)}{2\delta v}. \label{g(v)}%
\end{equation}
An additional argument for this choice is the fact that the same procedure
remains adequate when calculating the averaged $g(v)$ (see below). The
bias-voltage increment $\delta v$ was found to insignificantly modify the
result obtained if it is selected from the interval $0.0001\lesssim\delta
v\lesssim0.01$. Smaller $\delta v$-values gave rise to a \textquotedblleft
noise\textquotedblright\ associated with the finite accuracy of\ numerical
calculations, and larger ones to the redundant smoothing of CVC peculiarities,
especially noticeable at low temperatures. In specific calculations, we used
the value $\delta v=0.001$.

\begin{figure}[tb]
\includegraphics[width=\figww]{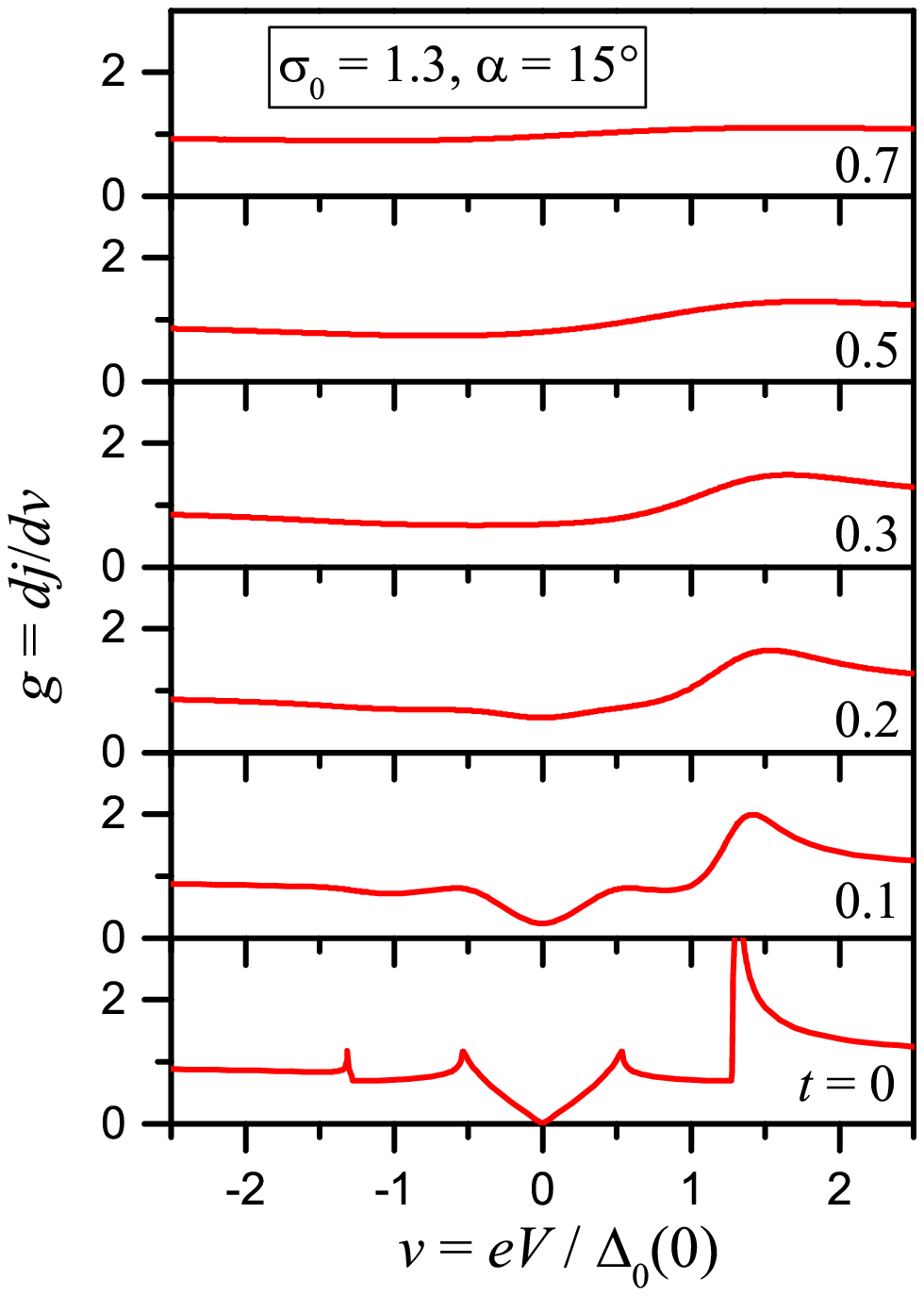}(a)
\includegraphics[width=\figww]{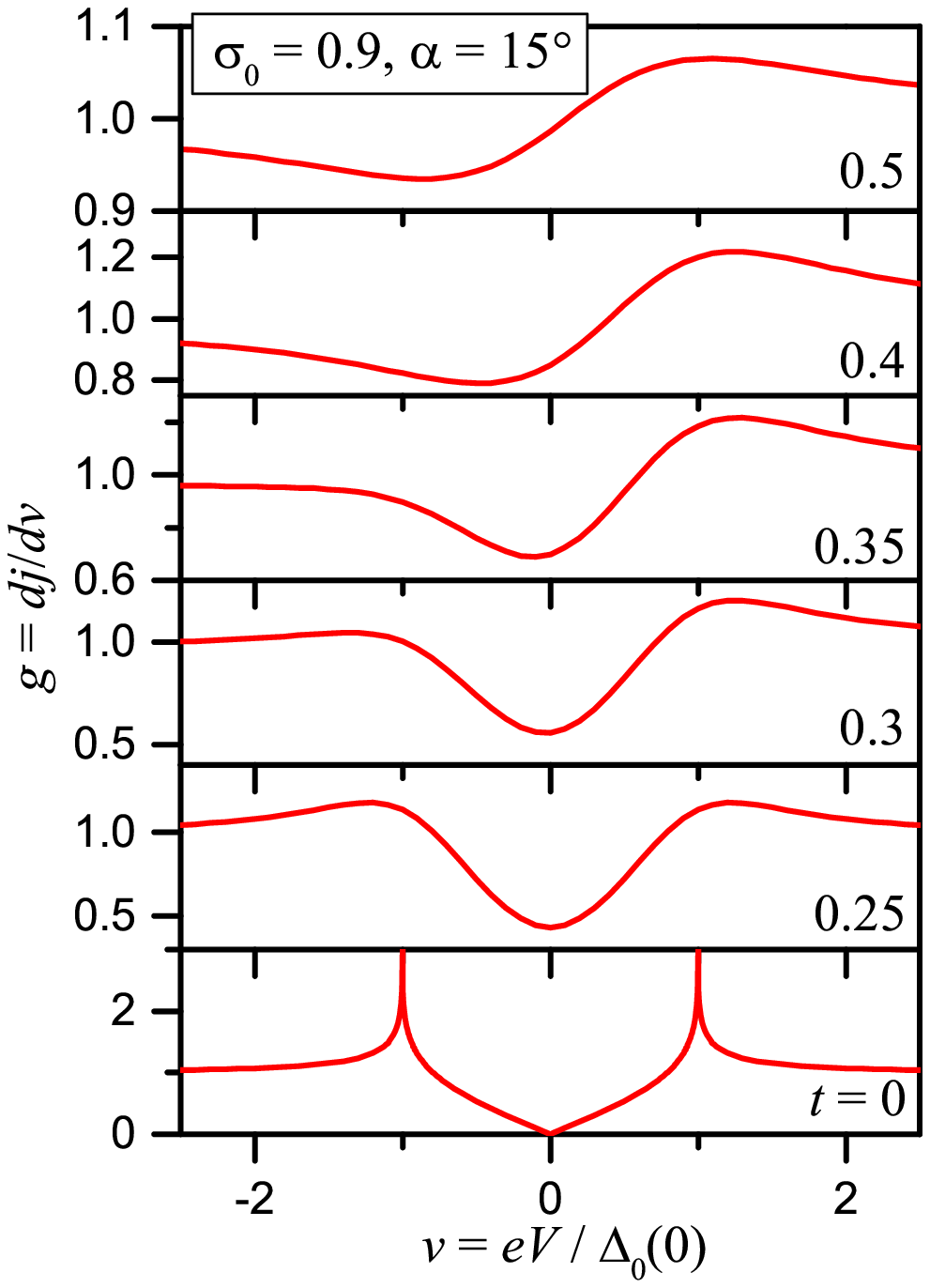}(b) \caption{(Color online)
Temperature evolution of the dependences of the dimensionless conductances
$g=R\frac{dJ}{dV}$ where $J$ is the quasiparticle current, $V$ the bias
voltage, and $R$ the resistance of the junction in the normal state on the
dimensionless voltage $v=eV/\Delta_{0}(0)$, for non-symmetric tunnel junction
between a partially dielectrized $d$-wave CDW superconductor (CDWS) and a
normal metal. $e>0$ is the elementary charge. Panels (a) and (b) correspond to
the regimes without and with the CDW reentrance. The parameters $\sigma_{0}$
and $\alpha$ are the same as in the corresponding panels in Fig.~\ref{fig1}.}%
\label{fig2}%
\end{figure}

In this article, we do not carry out a full analysis of the tunnel CVCs with
respect to the choice of problem parameters. Therefore, in Fig.~\ref{fig2}, we
present only the results of calculations representing the influence of $T$ on
the CVC shape for two characteristic CDWS cases, without [panel~(a)] and with
[panel~(b)] the $\Sigma$-reentrance. The panels demonstrate how effectively
CDWs distort the CVC symmetricity. It is especially clearly seen from
panel~(b), where the dimensionless reentrance temperature $t_{r}=T_{r}%
/\Delta_{0}(0)\approx0.28$, and the CVCs remain symmetric below this
temperature, because the CDWs are totally suppressed at $t<t_{r}$ [see
Fig.~1(b)]. In both cases, the structure of PG depletion above the critical
temperature [$t_{c}=T_{c}/\Delta_{0}(0)\approx0.23$ in panel~(a) and 0.41 in
panel~(b)] is reproduced excellently. But at low $T$'s, the CDW-induced peaks
[Fig.~2(a)] are much stronger than the smeared humps inherent to junctions
with Bi$_{2}$Sr$_{2}$CaCu$_{2}$O$_{8+\delta}$ or YBa$_{2}$Cu$_{3}$%
O$_{7-\delta}$. Nevertheless, in several intrinsic tunneling structures such
huge pseudogap peaks comparable to coexisting superconductivity-related
coherence peaks were also
observed,\cite{krasnov00:5860,krasnov02:140504,yurgens03:147005,suzuki12:214529}
which might be caused by specific properties of those very junctions (a high
degree of uniformity, which agrees well with our conjectures). Earlier data on
the YBa$_{2}$Cu$_{3}$O$_{7-\delta}$/Pb (in the normal state)
junction\cite{gurvitch89:1008} also showed the coherence peak and the hump of
comparable heights.

\begin{figure}[tb]
\includegraphics[width=\figw]{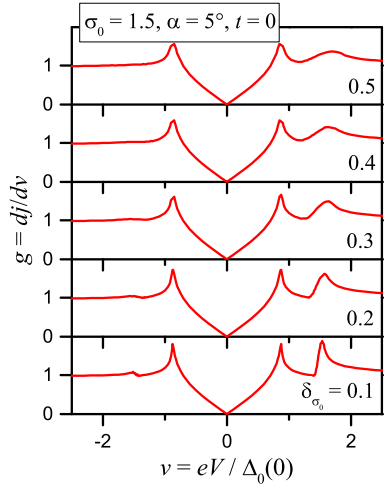} \caption{(Color online) Illustration
how the peak-dip-hump structure (PDHS) in $g(v)$\ at $t=0$ is formed under the
influence of the $\sigma_{0}$ spread, $\delta_{\sigma_{0}}$.}%
\label{figvarP}%
\end{figure}

Making allowance for the spatial non-uniformity of CDWS resolves this
discrepancy easily. In the simplest instance analyzed here, we assumed that
the non-uniformity reveals itself in the spread of parameter $\sigma_{0}$. The
calculations were carried out, similarly to
Ref.~\onlinecite{gabovich07:064516}, using formula~(\ref{g(v)}), where the
current $j$ was additionally averaged over the interval $[\sigma_{0}%
-\delta_{\sigma_{0}},\sigma_{0}+\delta_{\sigma_{0}}]$,
\begin{equation}
\left\langle j\right\rangle =\int_{\sigma_{0}-\delta_{\sigma_{0}}}^{\sigma
_{0}+\delta_{\sigma_{0}}}j(\sigma)w(\sigma)d\sigma%
%TCIMACRO{\TeXButton{TeX field}{\Big/}}%
%BeginExpansion
\Big/%
%EndExpansion
\int_{\sigma_{0}-\delta_{\sigma_{0}}}^{\sigma_{0}+\delta_{\sigma_{0}}}%
w(\sigma)d\sigma,
\end{equation}
with the bell-shaped weight function $w(\sigma)=((\sigma-\sigma_{0}%
)^{2}-\delta_{\sigma_{0}}^{2})^{2}$. Figure~\ref{figvarP} demonstrates the
effectiveness of this procedure that drastically smears the large CDW peak,
reduces its amplitude, and generates a PDHS typical of the CVCs for CDWS-I-N
junctions at low $T$'s.

\begin{figure}[tb]
\includegraphics[width=\figw]{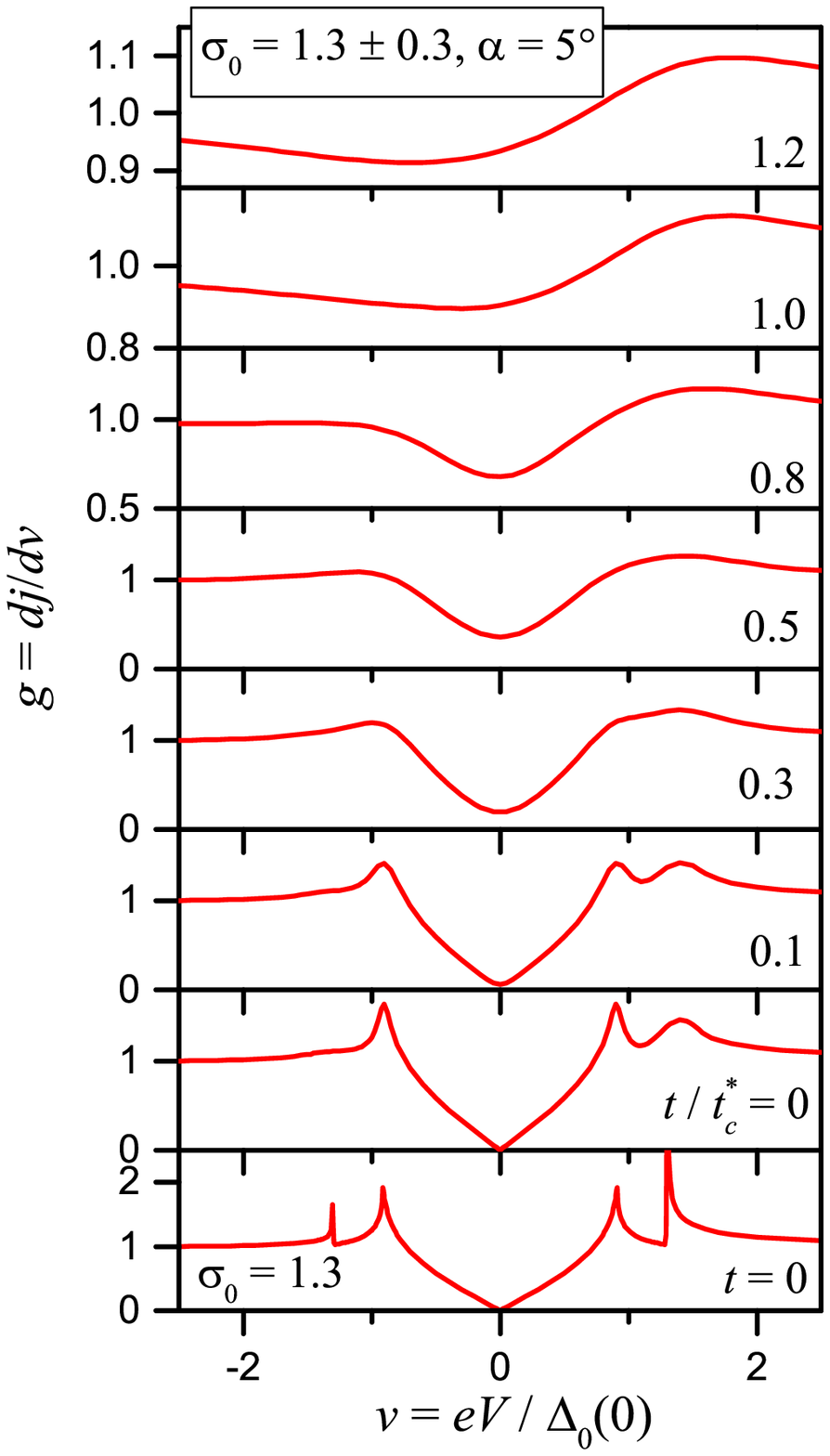} \caption{(Color online) Combined
action of temperature and $\sigma_{0}$ spread on $g(v)$. $t_{c}^{\ast}$ is the
critical temperature for the spatially uniform CDWS with $\sigma_{0}=1.3$ and
$\sigma=5^{\circ}$. The bottom panel corresponds to the CDWS with
$\delta_{\sigma_{0}}=0$.}%
\label{fig3}%
\end{figure}

In Fig.~\ref{fig3}, a combined action of the $\sigma_{0}$-spread and the
temperature is illustrated. The lowest two plots describe the formation of
PDHS. The other plots illustrate how all gross features in the observed CVC
transform into one another as $T$ grows. Specifically, at $T=0$, a noticeable
asymmetric PDHS is observed with the coherence peak higher than the hump in
the positive-bias branch. The negative-bias branch contains a coherence
superconducting peak and the remnants of the almost compensated and strongly
smeared CDW one. For $\varphi=\pi$, the branches would
interchange.\cite{gabovich07:064516} According to the accepted model it means
that CDWs are pinned with $\varphi=\pi$. As was said above, such a preference
cannot be explained in the framework of the phenomenological approach. In the
framework of our model, we can only guess that the applied electric field
rearranges the CDW superstructure near the CDWS-I interface to minimize the
total system energy. In this connection, it is important to bear in mind that
the CDW patterns at the high-$T_{c}$ oxide crystal surface are different from
their bulk counterparts, so that they may become vulnerable under the
influence of the applied Coulomb field.\cite{rosen13:1977}

Heating smears all gap-driven features, so that only a typical shallow
pseudogap depression remains, which is in agreement with the $c$-axis
intrinsic tunneling spectroscopy data for Bi$_{2}$Sr$_{2}$CaCu$_{2}%
$O$_{8+\delta}$.\cite{krasnov09:214510,suzuki12:010112} The depression extends
to relatively large $eV$ because the correction to Ohm's law is proportional
to
\begin{equation}
\frac{\Sigma_{0}(0)}{eV}\log\frac{eV}{\Sigma_{0}(0)}%
\end{equation}
due to the contribution from Green's function responsible for electron-hole
pairing.\cite{gabovich95:7437}

\begin{figure}[tb]
\includegraphics[width=\figww]{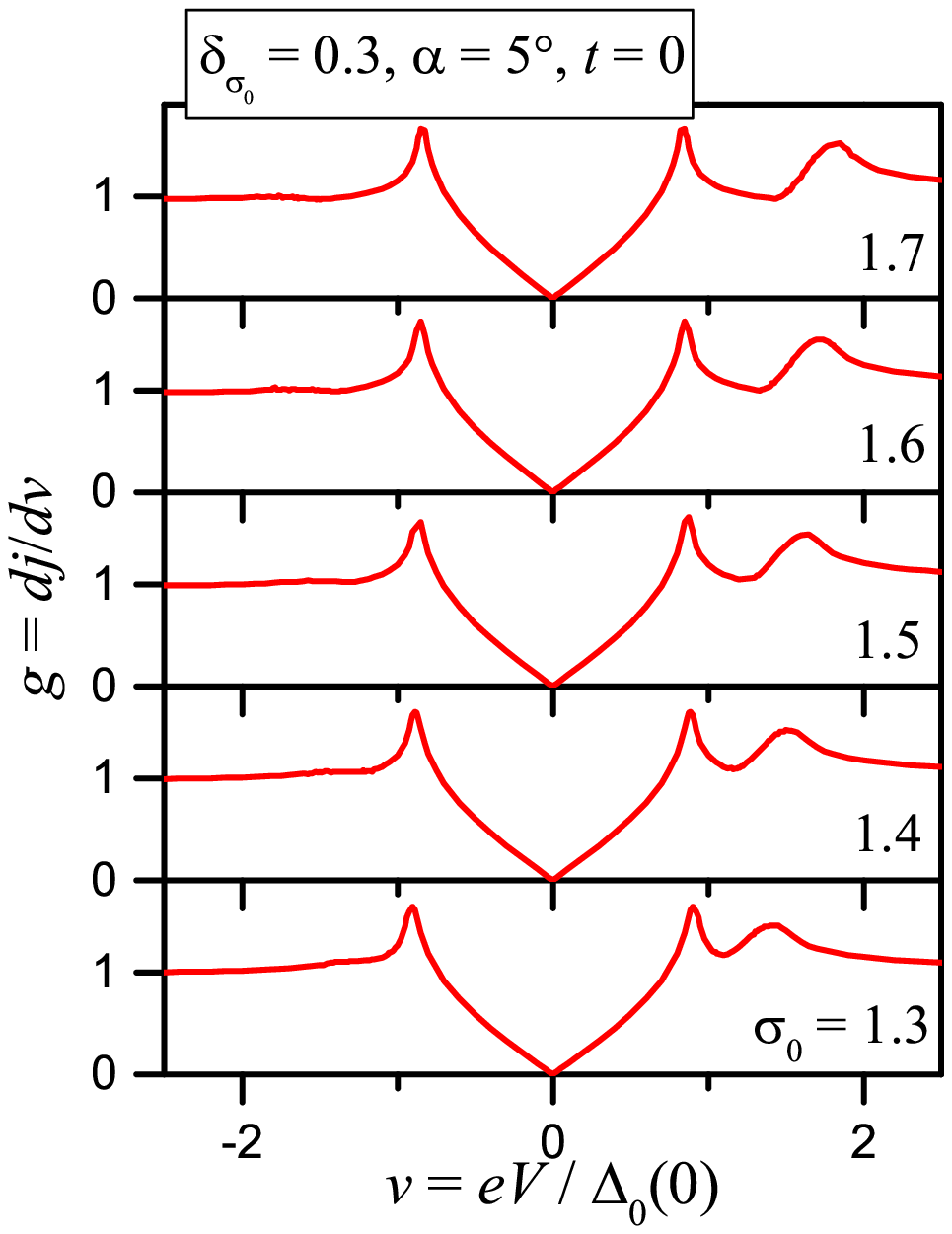}(a)
\includegraphics[width=\figww]{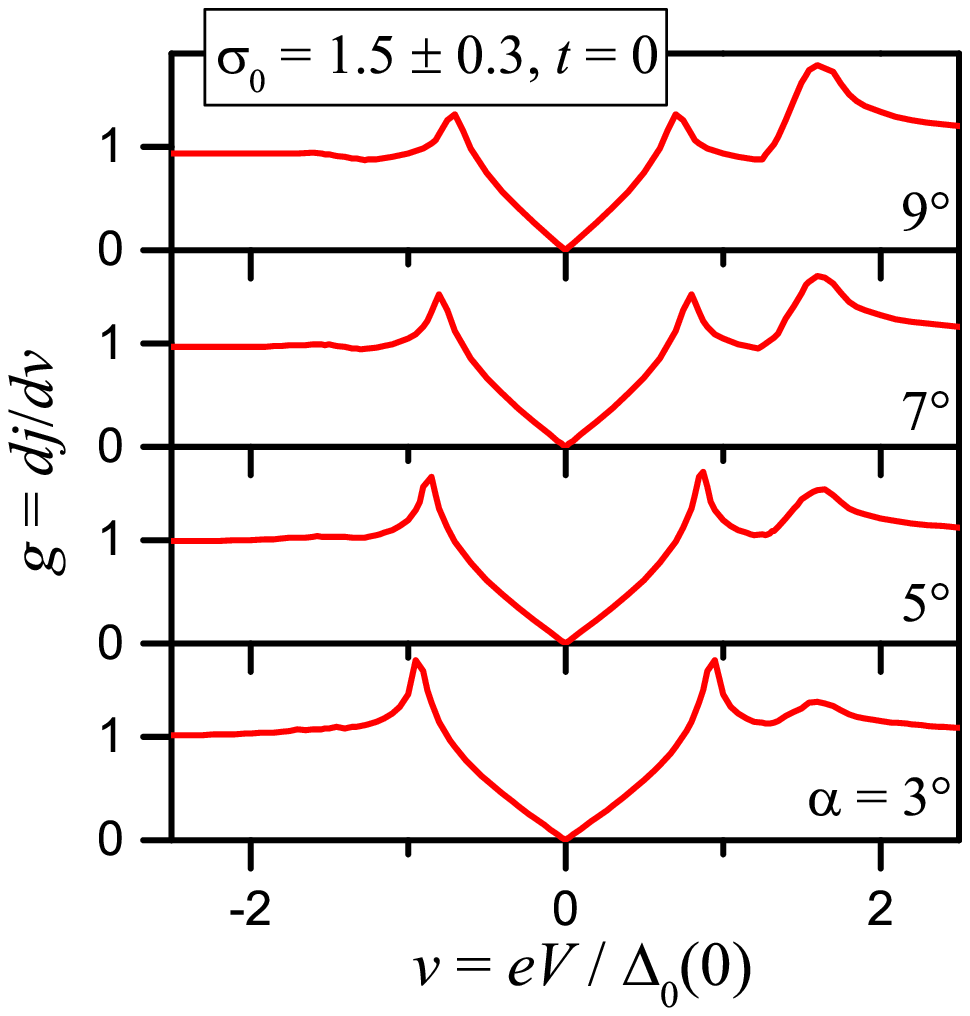}(b) \caption{(Color online) The
influence of the parameters $\sigma_{0}$ [panel (a)] and $\alpha$ [panel (b)]
on the PDHS.}%
\label{figVarSA}%
\end{figure}

It is worth emphasizing the following circumstance. As follows from
Figs.~\ref{figvarP} and~\ref{figVarSA}, by varying the parameters
$\delta_{\sigma_{0}}$, $\sigma_{0}$, and $\alpha$, we can substantially modify
the height of the smeared CDW hump and the distance between the hump and the
coherent peak in the PDHS. But the structures observed in real experiments are
reproduced the best when the parameter $\alpha$ falls within the interval
3--10$^{\circ}$. Estimations on the basis of experimental data bring about a
value of 15$^{\circ}$.\cite{gabovich13:104503} However, one should note that
our model contains certain simplifications. In particular, it assumes that
$\Sigma$ is constant, being the same on both pairs of nested FS sections and
vanishing in a jump-like manner at their boundaries; i.e. the influence of
CDWs on superconductivity is as strong as possible in the framework of the
assumptions made. More realistic corrections to those rather strict conditions
will inevitably enlarge the selected phenomenological value of $\alpha$ and
make it closer to the experimental one. It is necessary to bear in mind that
the Cooper pairing strength is inhomogeneous as well (since both involved
order parameters are
interrelated\cite{gabovich09:224501,gabovich10:681070,ekino11:699,ekino11:385701}%
), but to a much lesser extent.\cite{boyer07:802} That is why we considered
$\Delta_{0}(0)$ as a fixed normalizing parameter.

In order to find out the doping dependence of the tunnel spectra in the
framework of our model, it would have been necessary to change both control
parameters $\sigma_{0}$, and $\alpha$ simultaneously using the correlated
experimental data for certain cuprate families. However, to estimate the main
trend it is enough to change the parameter $\sigma_{0}$ alone. Then one can
see that the coherence-peak-to-hump energy distance increases nonlinearly with
$\sigma_{0}$, which is quite natural and corresponds to the transition from
the optimal-doping to underdoping compositions with larger PGs. It is in
accordance with the experiment, e.g., with tunnel data for Bi$_{2}$Sr$_{2}%
$Ca$_{2}$Cu$_{3}$O$_{10+\delta}$.\cite{berthod13:014528} We emphasize that the
original analysis of the same data based on the peak-to-dip energy dependence
seems to be misleading because the dip itself is an artifact as a trough
between two ridges, each having a certain physical meaning.

To summarize, we have shown that CDWs can be considered as the driving force
of the pseudogap gapping revealed in tunneling spectra of $d$-wave
superconductors. The spatial inhomogeneity transforms a smeared PG into a hump
which constitutes one shoulder of the PDHS observed below $T_{c}$. The
coherence peak of the superconducting origin forms another shoulder. The
resulting PDHS is observed only in one branch of the current-voltage
characteristics due to the loss of symmetry caused by the actual realization
of a certain CDW phase $\varphi$ ($0$ or $\pi$). Undoubtedly, the proposed
theory describing the role of CDWs in the formation of such ubiquitous
features in the CVCs of tunnel junctions involving high-$T_{c}$ oxides as the
pseudogap and the dip-hump structure is also applicable to the case of
symmetric junction and will be considered elsewhere.

The work was partially supported by the Project N~8 of the 2012-2014
Scientific Cooperation Agreement between Poland and Ukraine.

\end{document}